\documentclass[prb,twocolumn,showpacs,preprintnumbers]{revtex4}

\usepackage[dvipdfm]{graphicx}
\usepackage{amsmath}
\usepackage{amssymb}
\def\a{\alpha}
\def\b{\beta}
\def\D{\Delta}

\def\g{\gamma}
\def\k{\kappa}
\def\l{\lambda}
\def\s{\sigma}
\def\t{\tau}
\def\ua{\uparrow}
\def\da{\downarrow}

\begin{document}

\title{Quantum Monte Carlo study for multiorbital systems\\
with preserved spin and orbital rotational symmetries}

\author{Shiro Sakai,$^1$ Ryotaro Arita,$^{1,2*}$ Karsten Held,$^2$ 
and Hideo Aoki$^1$}

\affiliation{$^1$Department of Physics, University of Tokyo, Hongo,
Tokyo 113-0033, Japan\\
$^2$Max-Planck-Institut f\"{u}r Festk\"{o}rperforschung, Heisenbergstrasse\ 
1, Stuttgart 70569, Germany}

\date{\today}

\begin{abstract}
We propose to combine the Trotter decomposition and a series expansion 
of the partition function for Hund's exchange coupling 
in a quantum Monte Carlo (QMC) algorithm for  multiorbital
systems that preserves spin and orbital rotational symmetries. 
This enables us to treat the 
Hund's (spin-flip and pair-hopping) terms, which is difficult in 
the conventional QMC method.  
To demonstrate this, we first apply the algorithm to study 
ferromagnetism in the two-orbital Hubbard model
within the dynamical mean-field theory (DMFT).
The result reveals that the preservation 
of the SU(2) symmetry in Hund's exchange is important, where 
the Curie temperature is grossly overestimated when 
the symmetry is degraded, as is often done, to Ising (Z$_2$).  
We then calculate the $t_{2g}$ spectral functions of 
Sr$_2$RuO$_4$ by a three-band DMFT calculation 
with tight-binding parameters taken from the 
local density approximation with proper rotational symmetry.

\end{abstract}
\pacs{71.10.Fd; 71.15.-m; 71.27.+a}
\maketitle

\section{Introduction}
As was pointed out
by Slater as early as in the 1930s,\cite{s36}
the intra-atomic (Hund's) exchange 
interaction is of great importance for
ferromagnetism in transition metals.
Also 
the intensive study of transition-metal 
oxides---fueled by the discovery of high-temperature 
superconductivity---revealed the pivotal role of orbitals
and Hund's exchange, not only
for ferromagnetism but, e.g., also
for the  metal-insulator transition
\cite{if98,an02,l03,kk04,ik05,mg05,pb05,kk05,ah05} and
unconventional superconductivity.\cite{t00,h04,sa04}

For studying such correlation effects in multiorbital systems,
the dynamical mean-field theory\cite{mv89,gk96} (DMFT)
is one of the most successful methods. 
It takes into account  temporal fluctuations 
but neglects spatial ones,
and is known to describe the Mott-Hubbard transition.\cite{gk96}
Recently, the multiorbital Mott-Hubbard transition
has been studied most intensively by DMFT, 
\cite{pb05,an02,l03,kk04,ik05,mg05,kk05,ah05}
showing that the correct symmetry  of the exchange coupling
is essential, in particular, close to and at the Mott-Hubbard transition.
 
Also for
realistic material calculations,
multiorbital DMFT is important.
In particular, the so-called local density approximation (LDA)+DMFT 
method\cite{ap97}
allows for calculating strong correlations in materials
from first principles, and it has become one of the most active 
fields in solid state physics.\cite{ks05}
In this method, DMFT solves a model constructed from
 band structure obtained by the local density approximation.
Since materials such as transition-metal oxides 
usually have several orbitals around $E_F$,
the model is, in general, a multiorbital one.
This multiorbital Hubbard model hence consists of
the intra- and interorbital Coulomb interactions $U$ and $U'$,
Hund's-exchange and the pair-hopping interactions $J$,
as well as the one-electron hoppings obtained from the LDA.

DMFT maps the model self-consistently  
onto an effective impurity model
which includes the same on-site interactions 
on an impurity site and an infinite number of bath sites,
which are coupled to the impurity site through a hybridization
(hopping).
This impurity problem has to be solved numerically:
For multiorbital models, 
the exact diagonalization\cite{ck94} (ED)
and the Hirsch-Fye quantum Monte Carlo (HFQMC) method
\cite{hf85,gk96} are the  most widely employed nonperturbative impurity
solvers.
Both ED and HFQMC methods have their pros and cons and are hence
favorable in different situations.

ED
has the advantage that it can easily handle all the multiorbital 
interactions including Hund's coupling and the pair-hopping term.
However, it is severely restricted in the number of bath sites,
even  in the Lanczos
implementations 
which can take into account most bath sites
for $T=0$ or very low temperatures. \cite{cm05}
Since it cannot treat so many sites (and orbitals),
DMFT (ED) studies have been usually limited to two-orbital systems.

On the other hand, HFQMC simulation can handle not only two but more orbitals
and, in contrast to ED, also produces continuous spectra.
The former is important since there are various intriguing 
systems having 
three or more orbitals, especially $t_{2g}$ electron, 
perovskite-type transition-metal oxides,
exemplified by the spin-triplet superconductor.
${\rm Sr_2RuO_4}$\cite{mm03}
The continuous spectra are crucial for 
comparing with experiments, e.g., photoemission spectroscopy.
For these reasons the HFQMC method 
has been by far the most widely employed impurity solver 
as far as LDA+DMFT is concerned. 

In most DMFT HFQMC studies,\cite{remark1} however,
only the Ising ($z$) component  of
Hund's exchange coupling has been considered, with 
the pair-hopping term neglected.
The reason is that
the standard Hubbard-Stratonovich transformation\cite{h83}
is only applicable to the density-density part ($\hat{H}_U$) 
of the interactions and 
something else is required  for
the $x$ and $y$ components of Hund's coupling
and the pair-hopping interaction ($\hat{H}_J$).
A straightforward decoupling of $\hat{H}_J$
leads to a very severe sign problem,\cite{hv98}
making such QMC calculations virtually impossible.

In order to take into account $\hat{H}_J$,
we previously proposed a transformation with
a real and discrete auxiliary field
for these terms in two-orbital systems.\cite{sa04,h04}
So far, $s$-wave superconductivity\cite{sa04} and 
the orbital-selective Mott transition\cite{kk05,ah05} 
have been investigated with our transformation.

Despite these successes, some problems remain to be solved.
First, it is still difficult to treat more than two orbitals:
Since the exponential of $\hat{H}_J$ is not equal to 
the product of the exponentials of the distinct two-orbital parts
$\hat{H}_J^{mm'}$ ($m$ and $m'$ denote orbitals),\cite{remark2}
we cannot decouple  $\hat{H}_J$ for three or more orbitals 
by the simple product of two-orbital parts.
%
Second, the fermionic sign problem\cite{hv98} caused by $\hat{H}_J$
prevents us from studying  low temperatures.

In the present paper we propose and show that these difficulties are 
greatly remedied by employing 
a perturbation series expansion (PSE), 
separating out the problematic $\hat{H}_J$ term.
It should be noted that the method is numerically exact 
(and nonperturbative).  Also, the method is easily implemented, 
since our updating algorithm for sampling weights
and Green's function is basically the same as
the Hirsch-Fye algorithm, with 
the  differences being only for a factor multipling the weights
and an additional value for the auxiliary field.
While we focus here on DMFT, i.e., QMC simulations for impurity problems, 
the same ideas can be employed also for the lattice QMC method.

The auxiliary-field QMC method based on PSE for electron systems
was first proposed by Rombouts {\it et al.}\cite{rh99}
These authors applied PSE to the finite-size single-orbital Hubbard model 
and succeeded in obtaining results without time-discretization 
errors with less computational time than the conventional, 
Trotter-decomposition algorithm\cite.{bs81}
Although the method uses PSE, 
it counts all the contributions of the interaction
since the perturbation order taken into account is higher
than the maximum order of the samples above which
the weight is virtually zero. 
So the scheme is essentially nonperturbative.
Rubtsov {\it et al.} \cite{rl04} proposed another algorithm
to evaluate the series expansion of the partition function
and applied it to solve the DMFT equations. 
It does not involve any auxiliary fields but uses Wick's theorem.
Recently Werner {\it et al.} \cite{wc05} proposed to use 
a perturbation expansion starting from the strong-coupling limit.

In previous work,\cite{sa06} we extended Rombouts'
algorithm to the multiorbital Hubbard Hamiltonian,
using a similar transformation as in Ref.\ \onlinecite{sa04}  [Eq.\ (3)]
for $\hat{H}_J$, i.e.,
we expanded the Boltzmann factor (operator)  with respect to 
the total interaction $\hat{H}_U+\hat{H}_J$ shifted by a constant.
Then,
we  discretized the imaginary time $\beta=L\D\t$
and used a Hirsch-Fye-like updating algorithm
for solving the impurity model in the DMFT context.
Although the method significantly relaxes the sign problem
and allows, in principle, for more than two orbitals,
it turned out that the calculations are too heavy  
at low temperatures or for strong couplings,
especially for more than two orbitals.
That is because the computational effort increases with 
the perturbation orders appearing in the Monte Carlo samples.
This order can become very large (see Fig.\ 2) 
in multiorbital systems since there are many interactions:
$(2M-1)M$ terms in $\hat{H}_U$ and 
$2(M-1)M$ terms in $\hat{H}_J$ per site, 
where $M$ is the number of orbitals.

To overcome this difficulty, we here propose to combine 
the HFQMC and the PSEQMC methods, i.e.,
to adopt the series expansion for $\hat{H}_J$,
while the standard
Trotter decomposition and Hubbard-Stratonovich
decoupling are employed for $\hat{H}_U$. 
This algorithm
enables us not only to handle three or more orbitals
but also to reach much lower temperatures or stronger couplings than 
HFQMC (Ref.\ \onlinecite{sa04}) or PSEQMC calculations,\cite{sa06} 
even for the two-orbital Hubbard model.

In the following, we first introduce 
the multiorbital Hubbard model in Sec. \ref{Sec:model}
to turn to our algorithm in
Sec. \ref{Sec:method}, enclosing
details for the weight factor in the Appendix.
A validation and benchmarks are presented
in Sec. \ref{Sec:benchmarks}.  
Applications to local spin moments
and to ferromagnetism
in the two-orbital Hubbard model along with an application 
to Sr$_2$RuO$_4$ are discussed in
Sec. V.  
Section \ref{Sec:summary} summarizes the paper and gives an outlook.

\section{Model}
\label{Sec:model}
The $M$-orbital Hubbard Hamiltonian 
reads
\begin{eqnarray}\label{eq:hm}
  &\hat{H}& \equiv \hat{H}_0+\hat{H}_U+\hat{H}_J,
\label{Eq:Hubbard}
\end{eqnarray}
\begin{eqnarray*}
  &\hat{H}_0& \equiv -t\sum_{ij\s} \sum_{m=1}^M 
                 c_{im\s}^\dagger c_{jm\s} 
                -\mu\sum_{im\s}n_{im\s},\\
  &\hat{H}_U& \equiv 
       U \sum_{i,m}\left(n_{im\ua}n_{im\da}-\frac{n_{im\ua}+n_{im\da}}{2}\right)\\
    &&+U'\sum_{i,m<m',\s}\left(n_{im\s}n_{im'-\s}-\frac{n_{im\s}+n_{im'-\s}}{2}\right)\\
    &&+(U'-J)\sum_{i,m<m',\s}\left(n_{im\s}n_{im'\s}-\frac{n_{im\s}+n_{im'\s}}{2}\right),\\
      &\hat{H}_J& \equiv \sum_{m<m'}\hat{H}_J^{mm'}\\
  && \equiv J\sum_{i,m<m'}(
      c_{im\ua}^\dagger c_{im'\da}^\dagger c_{im\da} c_{im'\ua}\\
  && \hspace{45pt} 
     +c_{im\ua}^\dagger c_{im\da}^\dagger c_{im'\da}c_{im'\ua}+{\rm H.c.}),
\end{eqnarray*}
where $c_{im\s}^\dagger (c_{im\s})$ creates (annihilates) 
an electron with spin $\s$ in orbital $m$ at site $i$, 
and $n_{im\s}\equiv c_{im\s}^\dagger c_{im\s}$.
$\hat{H}_0$ describes the hopping of electrons 
between two neighboring sites $\langle ij \rangle$,
which we assume, unless otherwise indicated, to be orbital independent.
$\hat{H}_U$ is defined to be all the density-density interactions, i.e.,
the intraorbital ($U$) and interorbital ($U'$) Coulomb interactions 
and the $z$ (or Ising) component of Hund's coupling $J$.
By contrast, $\hat{H}_J$ takes care of the terms 
that cannot be written as  density-density interactions, which 
consist of the $x$ and $y$ component of 
Hund's coupling as well as the pair-hopping interaction 
(second term), in which two electrons in an orbital
transfer to other orbitals.
The Hamiltonian is rotationally invariant not only 
in spin space but also in real (orbital) space 
if we satisfy the condition $U\!=\!U'\!+\!2J$, 
which comes from the symmetry preserved between 
the Coulombic matrix elements for orbitals in a central field.  
We postulate the rotational invariance henceforth.

In the DMFT approximation, the Hamiltonian (\ref{Eq:Hubbard})
is approximated by an impurity problem
where the interactions $\hat{H}_U$ and 
$\hat{H}_J$ are restricted to a test atom (the impurity) 
embedded in the systems;
and $\hat{H}_0$ describes the coupling to
a noninteracting bath which has to be
determined self-consistently.\cite{remark3}
When we develop our algorithm in the next section, 
$\hat{H}_0$, $\hat{H}_J$, and  $\hat{H}_U$
denote these terms of the impurity
model, but the same ideas can also be applied
for lattice QMC simulations of the Hubbard model.

\section{(Trotter + Series-expansion) Method}
\label{Sec:method}
We start with the series expansion of the Boltzmann factor 
for continuous time after Ref.\ \onlinecite{rh99}.
However, here we 
perform this only for $\hat{H}_J$, i.e., 
 \begin{eqnarray}\label{eq:pse}
  &&e^{\g-\b\hat{H}}
   =e^{-\b (\hat{H}_0+\hat{H}_U)+\g -\b\hat{H}_J}\nonumber\\
 &=&e^{-\b (\hat{H}_0+\hat{H}_U)}+\sum_{k=1}^{\infty} 
    \int_0^1 dt_k \cdots \int_0^{t_2} dt_1 \prod_{i=1}^k\nonumber\\
 &&\left[ e^{-t_i\b (\hat{H}_0+\hat{H}_U)}(\g-\b\hat{H}_J)
    e^{t_i\b (\hat{H}_0+\hat{H}_U)} \right]
    e^{-\b (\hat{H}_0+\hat{H}_U)},\nonumber\\
 \end{eqnarray}
where we have shifted the Boltzmann factor by a constant
$\g$ for $\b\hat{H}_J$ 
to apply the auxiliary-field transformation Eq.\ (9) below.

Now we discretize the imaginary-time integrals
and with the notation $\hat{X}_1\equiv \g-\b\hat{H}_J$, 
Eq.\ (\ref{eq:pse}) equals
 \begin{eqnarray}\label{eq:dis}
 &&e^{-\b (\hat{H}_0+\hat{H}_U)}
  +\sum_{k=1}^{\infty} L^{-k}
   \sum_{j_k=1}^{L}\cdots\sum_{j_1=1}^{j_2}
   \prod_{i=1}^k \nonumber\\
 &&\left[ e^{-\frac{j_i}{L}\b (\hat{H}_0+\hat{H}_U)} \hat{X}_1
     e^{\frac{j_i}{L}\b (\hat{H}_0+\hat{H}_U)} \right]
     e^{-\b (\hat{H}_0+\hat{H}_U)}+O(\D\t).\nonumber\\
 \end{eqnarray}

We now show that 
this sum can be rewritten as
 \begin{align}\label{eq:rewrite}
  \sum_{s_1,\cdots,s_L}^{0,1} F(k;s_1,s_2,\cdots,s_L)
  \prod_{i=1}^L 
  [e^{-\D\t (\hat{H}_0+\hat{H}_U)}\hat{X}_{s_i}]\nonumber\\
  +O(\D\t),
 \end{align}
where $F$ is a positive weight factor, 
$k\equiv \sum_{i=1}^L s_i$, and $\hat{X}_0\equiv 1$.
To obtain the representation \ (\ref{eq:rewrite}),
we first cut off the $k$ summation in Eq.\ (\ref{eq:dis}) at $L$.
This cutoff is justified if $L$ is taken to be greater than 
the maximum perturbation order $k_{max}$ 
(defined and displayed below) 
appearing in the Monte Carlo samples, so that
there are no contributions from higher-order terms.
In practice, we can make $L$ much larger than 
$k_{max}$ (see Fig.2 below): $k_{max}$ depends on Hund's coupling $J$, 
where $J$ is physically not so large, whereas we can choose $L$ 
to satisfy $L>\b U$.

Second, 
we replace those terms having consecutive $\hat{X}_1$'s in 
Eq.\ (\ref{eq:dis}) by proximate terms including only 
one $\hat{X}_1$'s per imaginary time interval $\Delta \tau$. 
For example, 
$\cdots \hat{X}_1\hat{X}_1 e^{-\D\t (\hat{H}_0+\hat{H}_U)}\cdots$
is replaced by
$\cdots \hat{X}_1 e^{-\D\t (\hat{H}_0+\hat{H}_U)}\hat{X}_1\cdots$.
This replacement reduces the number of possible configurations
remarkably and casts the summation (\ref{eq:dis}) into the form
(\ref{eq:rewrite}) similar to the Trotter decomposition, which enables
us to employ their standard Hirsch-Fye algorithm with
only a slightly more complicated auxiliary field at each time
slice.
The error of this approximation (commutation)
is $O(\D\t)$, i.e., of the same order
as the time discretization,
as long as the average order of the series expansion
$\langle k \rangle$ is sufficiently smaller than $L$.
This is simply because the 
terms having two or more consecutive  $\hat{X}_1$'s 
rarely appear for $\langle k \rangle\ll L$.
For example, consider the second order terms in Eq.\ (\ref{eq:dis}).
Altogether, there are $L(L+1)/2$ second-order terms, but only $L$ 
of these terms have two consecutive $\hat{X}_1$'s with the same imaginary time 
interval. Hence, the error is at most $O(2\D\t/L)$.
Similar argument for higher orders justify the replacement as long
as $\langle k \rangle \ll L$.
Since we do not simply drop the terms with
two or more consecutive $\hat{X}_1$'s, but 
replace them by terms where the $\hat{X}_1$'s
are shifted to  neighboring imaginary time intervals,
we have to multiply the  Boltzmann factor by a factor $F$ to 
account for these replacements.
The detailed derivation of $F$ is given in the Appendix.

Now, we separate out $\hat{H}_U$ in Eq.\ (\ref{eq:rewrite}) 
using the Trotter decomposition as
 \begin{eqnarray}\label{eq:trot}
  e^{-\D\t (\hat{H}_0+\hat{H}_U)}=
  e^{-\D\t \hat{H}_0}e^{-\D\t\hat{H}_U}+O(\D\t^2),
 \end{eqnarray}
so that Eq.\ (\ref{eq:rewrite}) has a similar form to the
standard HFQMC method.
The $e^{-\D\t\hat{H}_U}$ term is then decoupled, as usual, 
into a sum of one-body exponentials with 
the Hubbard-Stratonovich transformation,
 \begin{align}\label{eq:HS}
   e^{-\D\t V [n_{\a}n_{\b}-\frac{n_{\a}+n_{\b}}{2}] }
   = \frac{1}{2}\sum_s^{\pm 1}\left\{ 
            \begin{array}{l}
            e^{\l_V s(n_{\a}-n_{\b})}
        \quad     (V\geq 0), \\
            e^{\l_V s(n_{\a}+n_{\b}-1)
           +\frac{a}{2}}  (V<0), 
            \end{array} \right.
 \end{align}
where $V$ stands for $U, U',$ or $U'-J$, 
$\l_V \equiv \ln (e^{|\D\t V|/2}
                        +\sqrt{e^{\left| \D\t V \right|}-1})$.  
We have also displayed the case of attractive interaction, 
which we shall require when we do the procedure described 
in Ref.\ \onlinecite{remark4}. 
Including
all the $(2M-1)M$ 
interactions of density-density type,
 the decoupling for $e^{-\D\t\hat{H}_U}$ is given by
\begin{eqnarray}\label{HSU}
  &&e^{-\D\t\hat{H}_U}
  =\sum_{P=1}^{N_U} \hat{Q}_P^U,\\
  &&\hat{Q}_P^U \equiv \frac{1}{N_U}
      \prod_{m=1}^M e^{ \l_U p_m(n_{m\ua}-n_{m\da})}\nonumber\\
  &&\times \prod_{m<m',\s}
       e^{ \l_{U'} q_{\s}^{mm'}(n_{m\s}-n_{m',-\s}) 
          +\l_{U'-J} r_{\s}^{mm'}(n_{m\s}-n_{m'\s}) },\nonumber
\end{eqnarray}
where $P (=1,...,N_U \equiv 2^{(2M-1)M})$
designates configurations of the auxiliary-field set
$( \{ p_m \},\{ q_{\s}^{mm'} \},\{ r_{\s}^{mm'} \} )$
with $p_m,q_{\s}^{mm'},r_{\s}^{mm'}=\pm 1$
denoting the fields for the $U$, $U'$, and $U'-J$ 
terms, respectively.

For $\hat{X}_1=\g-\b\hat{H}_J$ in Eq. (\ref{eq:rewrite}) 
we construct an auxiliary-field transformation 
as follows.\cite{sa06,rh98}
We first decompose $\hat{X}_1$ into the sum of 
all distinct two-orbital parts as
 \begin{align}\label{eq:HSJ1}
   \g-\b\hat{H}_J=\!
   \sum_{m<m'}[\g^{mm'}-\b\hat{H}_J^{mm'}],\quad
   \g\equiv\!\sum_{m<m'}\g^{mm'}.
 \end{align}
We then apply the decoupling
\begin{eqnarray}\label{eq:HSJ2}
  &&\g^{mm'}-\b\hat{H}_J^{mm'}=\nonumber\\
  &&\frac{\g^{mm'}-\b J}{8}\!\!\sum_{s,t_\ua,t_\da}^{\pm 1} 
  \prod_\s e^{\tilde{\l}_J[\s s f_\s^{mm'}+t_\s(n_{m\s}+n_{m'\s}-1)]}
  \nonumber\\
\end{eqnarray}
to every pair of orbitals, where
\begin{eqnarray*}
f_\s^{mm'}&\equiv& 
 c_{m\s}^\dagger c_{m'\s}+ c_{m'\s}^\dagger c_{m\s},\\
\tilde{\l}_J &\equiv& \frac{1}{2}\ln\frac{1+\k}{1-\k},\\
\k &\equiv& \sqrt{\frac{\b J}{\g^{mm'}}} < 1.
\end{eqnarray*}
Note that the difficulty in the HFQMC calculations
coming from the noncommutativity of $\hat{H}_J^{mm'}$'s 
is lifted in Eq.\ (\ref{eq:HSJ1}), so that 
we can readily deal with more than two orbitals.
Combining Eqs.\ (\ref{eq:HSJ1}) and (\ref{eq:HSJ2}), 
we end up with
 \begin{eqnarray}\label{eq:HSJ3}
 &&\g-\b\hat{H}_J=
   \sum_{S=1}^{N_J} \hat{Q}_S^J,\\
 &&\hat{Q}_S^J\equiv \frac{\g^{mm'}-\b J}{8}
   \prod_\s e^{ \tilde{\l}_J 
               [ \s s f_\s^{mm'}
                 +t_{\s}(n_{m\s}+n_{m'\s}-1)] },\nonumber 
 \end{eqnarray}
where $S [=1,...,N_J \equiv 4M(M-1)]$ 
corresponds to the set $(s,t_{\ua},t_{\da})$
for all the $M(M-1)/2$ pairs of 
orbitals $(m,m')$.

Collecting the addends from the 
decoupled $\hat{H}_U$ and  $\hat{H}_J$ terms,
we finally obtain 
 \begin{align}\label{eq:final}
  e^{\g-\b\hat{H}}=
  \sum_{S_1,\cdots,S_L}^{0,\cdots,N_J} 
  F(k;\tilde{s}_1,\tilde{s}_2,...,\tilde{s}_L)\nonumber\\
  \times\sum_{P_1,\cdots,P_L}^{1,\cdots,N_U}\prod_{i=1}^L 
  e^{-\D\t\hat{H}_0} \hat{Q}_{P_i}^U \hat{Q}_{S_i}^J+O(\D\t),
 \end{align}
with $\hat{Q}_0^J\equiv 1$
and
$\tilde{s_i}\equiv 0$ (for $S_i=0$) or 1 (otherwise).\cite{remark4}
Note that because $F(0;0,...,0)=1$, 
the zeroth-order term in Eq.\ (\ref{eq:final}) reproduces
the Hirsch-Fye algorithm with Ising-type Hund's coupling.

Owing to the form of Eq.\ (\ref{eq:final}), 
we can apply the same algorithm 
as in HFQMC for Monte Carlo sampling. Even the updating
equations for single spin flips are the same.

\section{Benchmarks}
\label{Sec:benchmarks}
\begin{figure}[t]
\includegraphics[width=7cm, height=5.5cm]{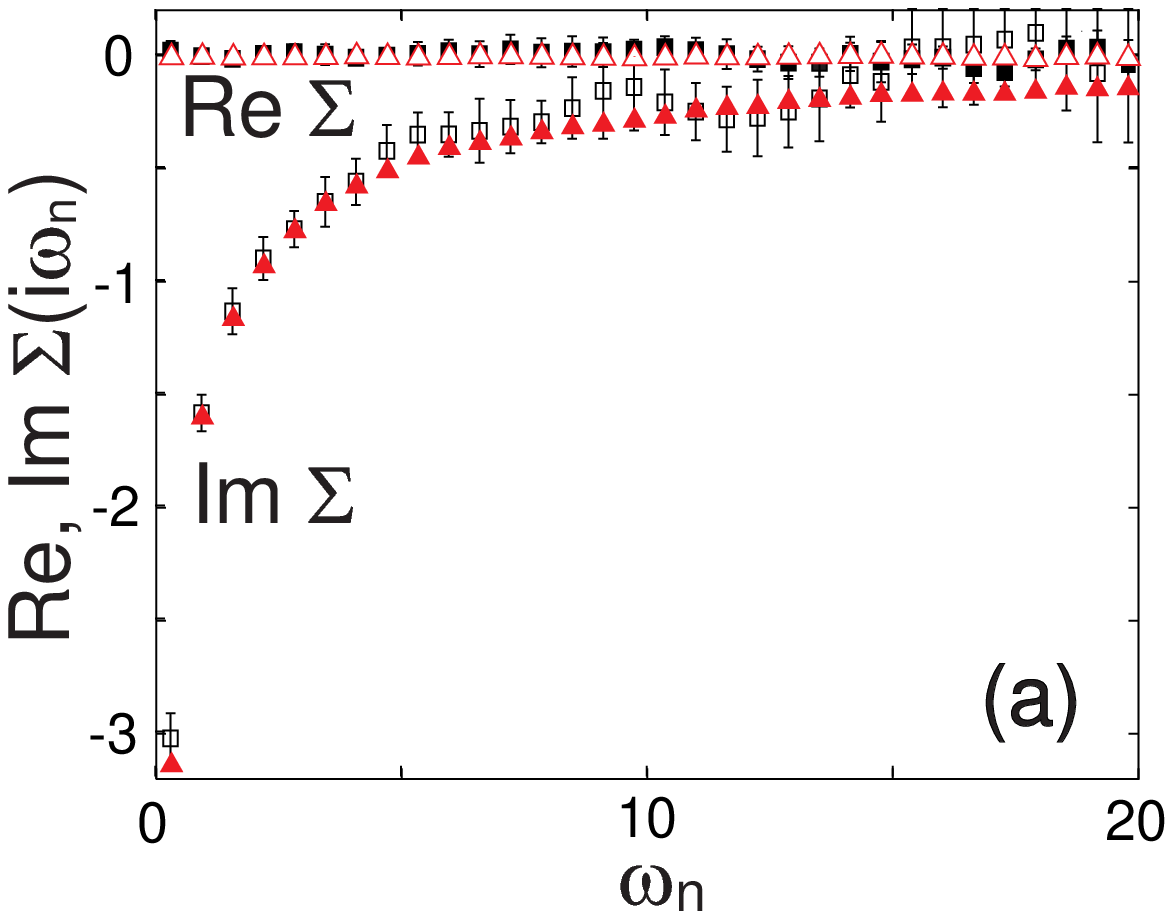}
\includegraphics[width=7cm, height=5.5cm]{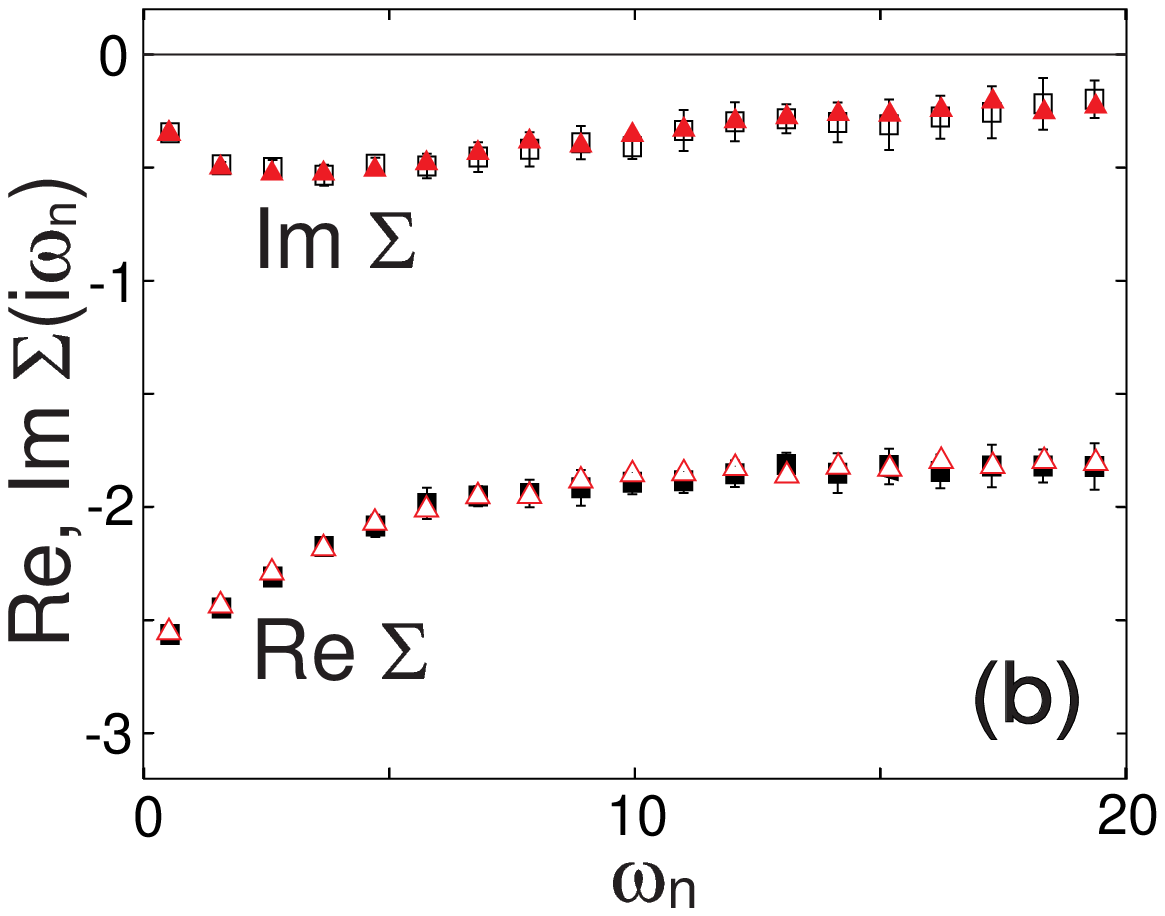}
\caption{(Color online) Real and imaginary parts of the self-energy
 against the Matsubara frequency $\omega_n$
for the two-orbital Hubbard model 
for (a) an insulating case with $n=2, \b=10, U'=2, J=0.4$, and 
(b) a metallic case with $n=1, \b=6, U'=2, J=1$.  
Result with the Hirsch-Fye algorithm (Ref.\ \onlinecite{sa04}) 
is shown in black squares, and the present QMC result in red triangles.
}
\label{fig:sig}
\end{figure}

As a benchmark, we compare in Fig.\ \ref{fig:sig} the electron 
self-energy obtained from our algorithm with that from the HFQMC method
of Ref.\ \onlinecite{sa04} for the two-orbital Hubbard model. 
In these QMC methods the self-energy is obtained from Dyson's
equation, where Green's function is calculated by averaging
over QMC samples.
We chose the hypercubic lattice with bandwidth $W=2$, and took 
$6\times 10^6$ Monte Carlo samples for both methods. 
We can see that the two results agree with each other 
within error bars for both
(a) an insulating case at half filling $n=2$ with $\b=10,U'=2,J=0.4,L=100$, 
and 
(b) a metallic case at $n=1,\b=6,U'=2,J=1,L=64$.
We notice, however, that 
the statistical error is much smaller in the 
present QMC approach than in the HFQMC one.
This is because the number of negative signs is greatly
reduced in the present scheme: 
The sign problem is mitigated.
Quantitatively, the average sign in the QMC weights is 
0.01 (0.03) for HFQMC methods while they are increased to 
0.30 (0.50) in the present algorithm
in case (a) [(b)].
%
%
This also implies that the present method can reach much lower temperatures.
We note that, while $\gamma$ is arbitrary, we adopted in these and following 
calculations $\gamma-\beta V \sim 0.1-0.3$, which has turned out to 
reduce the computational time to some extent.

\begin{figure}[t]
\includegraphics[width=6.5cm, height=5cm]{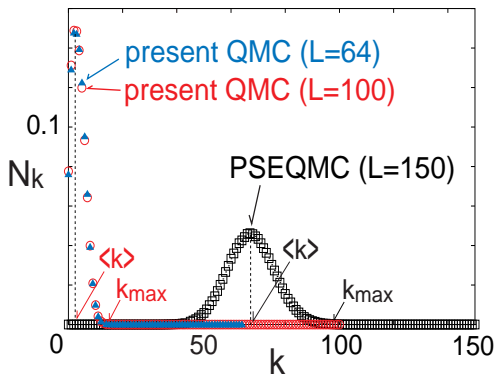}
\caption{(Color online) 
The distribution $N_k$ of the order of perturbation $k$
for the two-orbital Hubbard model with $n=1.9,\b=8,U'=4,J=0.2$
obtained with the present algorithm ($L=64$ or $100$) and 
with the PSEQMC algorithm ($L=150$) (Ref.\ \onlinecite{sa06}).}
\label{fig:nk}	
\end{figure}

Figure \ \ref{fig:nk} depicts a typical distribution $N_k$ of the 
order of perturbation 
$k$ contributing in the Monte Carlo simulation 
for $n=1.9,\b=8,U'=4,J=0.2$.
We can immediately see that the present algorithm has 
a peak in the distribution residing at 
much lower $k$ than for the PSEQMC approach.\cite{sa06}
This is natural, 
since the present method uses the expansion 
only with respect to $\hat{H}_J$, 
while the PSEQMC method expands with respect to the total 
interaction $\hat{H}_U+\hat{H}_J$. 
We notice that the weight is virtually zero above 
a certain order, $k_{max}$, in actual calculations, 
so that, although the method is called perturbation expansion, 
it takes account of all orders in fact.  
We find that the maximum order in the distribution 
is $k_{max}\sim 100$ for the PSEQMC method, 
while $k_{max}\sim 15$ is much lowered for the present QMC method. 
This means that $L$ must be taken to be $>100$ for the PSEQMC method
to take care of all orders, 
while $L>\b U \sim 35$ suffices for the 
present algorithm to take care of all orders. 
Such a smaller value of $L$ dramatically reduces the
computational effort in QMC simulations, which increases 
proportionately to $L^3$.
Moreover, the average order $\langle k \rangle$ 
is about 4 for the present QMC, 
which means that the approximation employed to obtain 
the form (\ref{eq:rewrite}) has only a very minor effect  
on the results, and hence is justifiable.
This can also be confirmed from the fact that the results do 
not significantly depend on $L$. 

\begin{figure}[t]
\includegraphics[width=6.5cm, height=5cm]{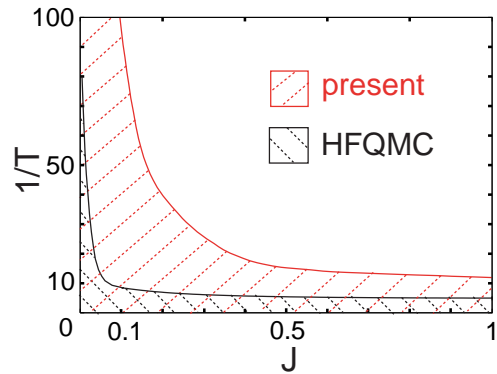}
\caption{(Color online) 
Computable regions (hatched) in the $T$-$J$ parameter space 
that can be computed  with the present and with
the Hirsch-Fye QMC method (Ref.\ \onlinecite{sa04}) for the 
two-orbital Hubbard model with $U'=4,W=2$.
Here we define the computability by
requiring average signs to be greater than 0.01.}
\label{fig:norm}
\end{figure}

Figure \ref{fig:norm} shows the computable regions 
for the present QMC and 
for the HFQMC methods (Ref.\ \onlinecite{sa04}) when $\hat{H}_J$ is included.
Here we define the region as computable when 
the average sign is greater than 0.01.
We can see that a much wider parameter region becomes 
computable in the present algorithm than in 
the HFQMC method. For small $J\ (\lesssim 0.2)$, we can explore
five to ten times lower temperatures. 
We attribute this improvement to the fact that 
$\hat{H}_J$ (which is the source of negative weights)
appears $L$ times for every sample in the HFQMC method, 
while we have only $\langle k \rangle$ such
terms on average in the present QMC algorithm.

\section{Applications}
\label{Sec:applications}

\subsection{Local spin moment}
\label{Sec:localspin}
\label{Sec:localmoment}

\begin{figure}[t]
\includegraphics[width=6.5cm, height=4.5cm]{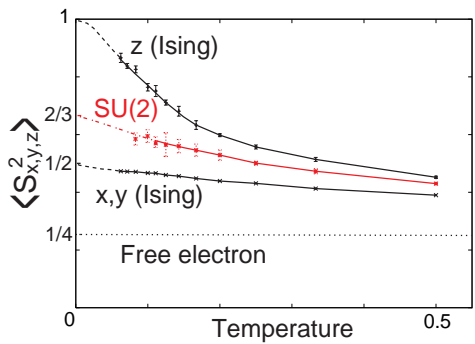}
\caption{(Color online) Local spin moment for the two-orbital
 Hubbard model with SU(2)-type Hund's and pair-hopping interactions 
 as compared with Ising-type Hund's coupling 
 for $U'=3, J=0.2, n=2$ (half filling) 
 and the Gaussian density of states of width $W=2$.  
 Solid lines are guides to the eye, 
 dashed lines extrapolations, and 
the dotted line the local moment for free electrons.}
\label{fig:moment}
\end{figure}

Let us now turn to the first application of our algorithm. 
We first examine how results for the Hamiltonian 
with and without $\hat{H}_J$, i.e.,
with Z$_2$ and SU(2) symmetry, respectively, can differ.
In Fig.\ \ref{fig:moment},
we show the temperature dependence of 
the local spin moment in the paramagnetic phase 
at half filling in the two-orbital Hubbard model 
for $U'=3, J=0.2$ on 
a hypercubic lattice with bandwidth $W=2$.
At these parameters,
 the system becomes insulating
at zero temperature for both the Z$_2$\ (Ising) and the
SU(2) cases.
The spin moments are defined by
\begin{eqnarray}
 &\langle S_z^2\rangle&
=\frac{1}{4}\langle (\sum_{ms}\s n_{m\s})^2 \rangle,\\
 &\langle S_x^2\rangle&
=\frac{1}{4}\langle (\sum_{m} c_{m\ua}^\dagger c_{m\da}
                     +c_{m\da}^\dagger c_{m\ua})^2 \rangle,\\
 &\langle S_y^2\rangle&
=-\frac{1}{4}\langle (\sum_{m} c_{m\ua}^\dagger c_{m\da}
                      -c_{m\da}^\dagger c_{m\ua})^2 \rangle
=\langle S_x^2\rangle.
\end{eqnarray}

For the Ising case $\langle S_z^2\rangle$ is always 
larger than $\langle S_{x,y}^2\rangle$,
 while 
for the SU(2) case these components have 
the same value (within statistical error bars) as they should, 
which lies between 
the $\langle S_z\rangle$ and $\langle S_x\rangle, 
\langle S_y\rangle$ in the Ising case.
As the temperature is lowered in the Ising case,
$\langle S_z^2\rangle$ and $\langle S_{x,y}^2\rangle$ 
 approach 1 and $\frac{1}{2}$, respectively,
while for SU(2) symmetry  both expectation values
converge to $\frac{2}{3}$.
This is easily understood by considering 
the fact that the ground state in the atomic limit
is a doublet with $S_z=\pm 1$ 
in the Z$_2$ case,
 whereas it is a triplet  with $S_z=0,\pm 1$ in the
SU(2) case.

\subsection{Ferromagnetism}
\label{Sec:ferromagnetism}

\begin{figure}[t]
\includegraphics[width=6cm, height=4.5cm]{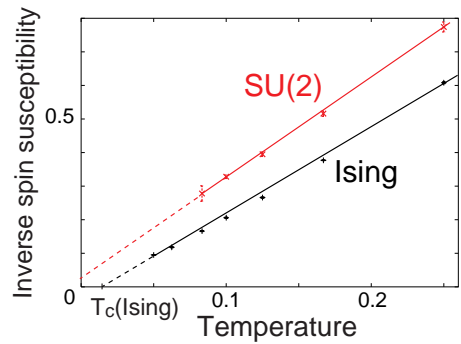}
\caption{(Color online) Inverse spin susceptibility for the two-orbital
 Hubbard model with SU(2)-type Hund's and pair-hopping interactions 
 as compared with the Ising-type Hund's coupling 
 for $U'=2.5, J=1$ with the semielliptical density of states 
 of width $W=2$.  
 The solid lines are guides to the eye, and
 the dashed lines extrapolations.}
\label{fig:susc}
\end{figure}

We now turn to the long-range ferromagnetic ordering,
which reveals a more important difference between Z$_2$ and SU(2) symmetries
as  Fig.\ \ref{fig:susc} shows.
To this end, we have calculated the temperature dependence of 
the magnetic susceptibility in the two-orbital Hubbard model
for $U'=2.5, J=1, n=1.25$ 
with a semielliptical density of states of width $W=2$.
The lattice susceptibilities are obtained here through
the Bethe-Salpeter equation.\cite{gk96}
The Curie temperature for the Hamiltonian with 
Ising-type Hund's coupling is estimated by a Curie-Weiss
extrapolation to be  0.01-0.02, 
in accordance with the previous results.\cite{hv98}
In contrast, the present result shows that 
the inverse susceptibility does not intersect the 
horizontal axis, implying an absence of ferromagnetic transitions, 
for these parameters if the SU(2) symmetry of
Hund's exchange and the pair-hopping term are respected.

The result suggests that an Ising-type treatment of Hund's coupling
grossly overestimates the tendencies
toward ferromagnetic ordering.
Indeed, Ising-type  DMFT calculations for manganites \cite{hv00}
and iron \cite{lk01} gave Curie temperatures
much higher than the experimental results.  
The present comparison indicates that the main reason for 
overestimating the Curie temperature in such calculations
is not only the mean-field nature of DMFT but 
the incorrect symmetry of the Hund's exchange.
For an Ising-type Hund's exchange, 
(local) transverse spin fluctuations, which are the main source 
of softening magnetic moments, are suppressed.

\subsection{Quasiparticle density of states for ${\rm Sr_2RuO_4}$}
\label{Sec:Sr2RuO4}

\begin{figure}[t]
\includegraphics[width=6cm, height=4.5cm]{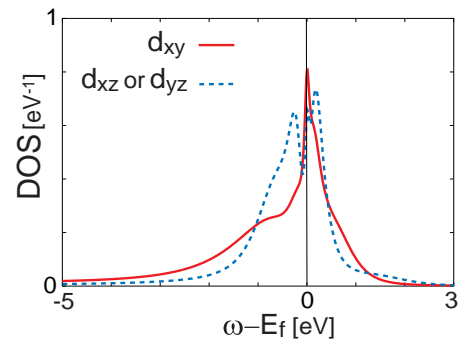}
\caption{(Color online) Quasiparticle density of states for $t_{2g}$ orbitals
in ${\rm Sr_2RuO_4}$, obtained by a simplified LDA+DMFT\ (QMC) 
calculation with three-orbital rotationally invariant 
Hamiltonian (\ref{eq:hm}).}
\label{fig:SRO}
\end{figure}

Finally we demonstrate that the present QMC method is actually 
applicable to realistic three-orbital systems.
As an illustration, 
we take ${\rm Sr_2RuO_4}$, 
which can be treated as a ($t_{2g}$) three-orbital system, 
consisting of a wide two-dimensional band ($d_{xy}$) and 
two degenerate narrow one-dimensional bands ($d_{xz},d_{yz}$).
The material belongs to an important family of transition-metal oxides where 
superconductivity,\cite{mm03} magnetism,\cite{cm66} 
and orbital-selective Mott transition\cite{an02,l03,kk04,ik05,mg05}
are discussed. 
To our knowledge, this is the first QMC calculation for
a three-orbital system with SU(2)-symmetric Hund's exchange
and pair-hopping interaction.

Liebsch and Lichtenstein \cite{ll00} already 
obtained the quasiparticle density of states for this material 
with a simplified LDA+DMFT\ (HFQMC) calculation using
the three-orbital Hamiltonian without $\hat{H}_J$ 
(i.e., an Ising-type treatment of Hund's coupling).
As a LDA input they constructed a tight-binding model 
that reproduces the LDA density of states.
We have employed the same initial tight-binding model 
(for the density of states), the same
number of $t_{2g}$ electrons per site ($n=3.7$), and the
same interaction parameters
$U'=0.8\,$eV, $J=0.2\,$eV, but a somewhat higher temperature 
$T=25\,$meV.

The LDA+DMFT spectral functions, obtained with the maximum 
entropy method, are shown in Fig.\ \ref{fig:SRO}.
The quasi-two-dimensional\ (2D) $d_{xy}$ band has a van Hove singularity 
just above the Fermi energy $E_F$,
while the quasi-1D $d_{xz}$ and $d_{yz}$ bands have broad peaks
on both sides of $E_F$ and 
a small peak near $E_F$.
Although these structures are similar to those
of Liebsch and Lichtenstein, \cite{ll00} 
it may be due to the small value of $J$ employed here.
In general, $\hat{H}_J$ can have a significant influence
on quasiparticle spectra.
We have obtained results substantially changed 
by $\hat{H}_J$, for example, enhancement of Kondo 
resonance by $\hat{H}_J$, 
for artificially elevated $U$,\ $U'$, and $J$
(not shown here).
Actually, values of $U$,\ $U'$, and $J$ larger than
those used here, 
as usually assumed for transition-metal oxides,
were recently suggested for Sr$_2$RuO$_4$,\cite{pn06}
and it will enhance the effect of $\hat{H}_J$.

\section{Summary and Outlook}
\label{Sec:summary}

In summary, we have formulated a numerically exact auxiliary-field 
QMC scheme for the spin- and orbital-rotationally-invariant 
Hamiltonian (\ref{eq:hm}) 
by combining series expansion and 
Trotter decomposition.
We have implemented this algorithm for 
impurity models and employed it
to solve the DMFT equations.
The approach enables us not only to address three-\ (or more-)orbital 
systems but also to reach much lower temperatures than the previous 
QMC methods, which is the case with two-orbital systems as well.

The calculation of the magnetic susceptibility shows
that an Ising treatment of Hund's coupling overestimates
tendencies towards ferromagnetism,  which we attribute
to neglected transversal spin fluctuations.
In this sense the SU(2) symmetry in  Hund's exchange 
is indicated to be mandatory for an accurate study of
magnetic phenomena with (LDA+)DMFT.
We finally applied the present method to a realistic three-orbital system
${\rm Sr_2RuO_4}$,
to demonstrate that the QMC algorithm
can be employed for LDA+DMFT calculations.

Since we can apply the usual Hirsch-Fye updating algorithm
with only minor changes in the additional 
values of the auxiliary field and a factor for the weights, 
the present scheme can be conveniently applied to 
multiorbital DMFT studies, especially for LDA+DMFT
 calculations requiring three orbitals.
Although the fermionic sign problem still remains 
at low temperatures, we can reach much lower
temperatures than before.\cite{sa04}
Also,  a combination of the present method
with the projective QMC method\cite{fh04,ah05}
is very promising, which may provide an avenue for examining 
ground states, which is now under way.

\section*{Acknowledgments}
We wish to acknowledge fruitful discussions with A. Muramatsu, T. Oka, 
and particularly Y.-F. Yang.
Numerical calculations were performed at the facilities
of the Supercomputer Center, Institute for Solid State Physics,
University of Tokyo and at the Information Technology Center, 
University of Tokyo.
Financial supports from the Alexander von
Humboldt Foundation (R.A.), the
Deutsche Forschungsgemeinschaft through the Emmy Noether program
(K.H.), and
the Japanese Ministry of Education, Culture, Sports, 
Science and Technology through Special Coordination
Funds for Promoting Science and Technology
are acknowledged.

\section*{Appendix}
We derive here the 
factor $F(k;s_1,s_2,...,s_L)$
$(s_i=0,1;k=\sum_{i=1}^L s_i)$ for Eq.\ (\ref{eq:rewrite}).
We introduce this factor to account for the 
contribution from the terms with consecutive $\hat{X}_1$'s 
at the same imaginary-time interval
in the sum (\ref{eq:dis}).
These terms
have been replaced with terms
having $\hat{X}_1$'s on proximate imaginary-time
intervals in  Eq.\ (\ref{eq:rewrite}).
In the following we abbreviate 
$e^{-\D\t(\hat{H}_0+\hat{H}_U)}$ as $h$, and $\hat{X}_1$ as $x$.

Central to our considerations are those
terms that include a substring
$xhxhx\cdots hx$
where $x$ and $h$ appear alternately
$m$ and $m-1$ times, respectively, with $2\leq m \leq L$.
This is because any term with
consecutive $x$'s will be replaced
with a term having such substrings.
For example, 
$xxhhx\cdots hx$ and $xhxxh\cdots hx$
are both approximately the same as $xhxhx...hx$
[where commuting an $x$ and
 a $h$ yields an error $\sim O(\Delta \tau)$].
In general, we commute $x$'s and $h$'s until there
are no consecutive $x$'s any longer.
Hence, we end up with an 
alternation of $x$'s and $h$'s, i.e.,
an $xhxhx...hx$ substring.
Because of these replacements, terms having such
a substring have to be weighted more.
In the following,  we construct a
rule for the replacement and weighting factor, 
avoiding a double counting.

Let us denote by $i$ a position
(from the left) in 
a substring which consists 
of $m$ $x$'s and $(m-1)$ $h$'s altogether.
We define $\a_i$ and $\b_i$ as the number of $x$'s and $h$'s
that are in $[1,i]$.
All substrings having 
\begin{eqnarray}\label{eq:ap1}
  \a_i \geq \b_i \quad {\rm for\ all}\ i
\end{eqnarray}
will be replaced 
with $xhxhx...hx$ which has alternating $x$'s and $h$'s.

For example, for  $m=2$
$xhx$ and $xxh$ will be replaced
with
$xhx$; for $m=3$,
$xhxhx,xhxxh,xxhhx,xxhxh$ and $xxxhh$
will all be replaced with $xhxhx$.
The condition (\ref{eq:ap1}) 
is necessary
to avoid a double counting.
However, the condition (\ref{eq:ap1}) excludes 
the substrings situated at the end of the imaginary time 
interval for which $\exists\ i'$, so that $ \a_{i'}<\b_{i'}$.
These terms are replaced with a substring $xhxhx\cdots hx$, 
where the last $x$ is at $L$,
namely, only when the last $x$ is at $L$,
does $xhxhx\cdots hx$ replace 
all the substrings having $m$ $x$'s and $(m-1)$ $h$'s,
which requires a separate treatment ($c$ factors below).

A second factor to be taken into account is 
the correction of the volume in the imaginary-time integrals;
namely, the weight for those terms having consecutive
$x$'s at
\begin{eqnarray}\label{eq:ap2}
 j_i\neq j_{i+1}=j_{i+2}=\cdots =j_{i+l}\neq j_{i+l+1}
\end{eqnarray}
in the sum (\ref{eq:dis}) 
should be reduced by a factor $\frac{1}{l!}$,
since the imaginary times originally satisfy a relation
\begin{eqnarray}\label{eq:ap3}
     t_{i+1} < t_{i+2} < \cdots < t_{i+l} 
\end{eqnarray}
in Eq.\ (\ref{eq:pse}). Hence the volume for 
$l$ consecutive $x$ terms, as in Eq.\ (\ref{eq:ap2}),
should be reduced to $\frac{L^{-l}}{l!}$.

Let us now introduce the quantity $b(i,j)$ 
for the weight (apart from the volume factor $L^{-j}$) 
of all the substrings having $i$ $h$'s and $j$ $x$'s.
For $j$ consecutive $x$'s, we simply 
have the aforementioned  $1/l!$ factor, i.e.,
\begin{eqnarray}
  b(0,j)=\frac{1}{j!} \quad {\rm for} \quad 0 \leq j \leq L.
\end{eqnarray}
This is the starting point for the  recurrence
 formula, 
\begin{eqnarray}\label{eq:b}
  b(i,j)=\sum_{k=i}^j 
         \frac{1}{(j-k)!}b(i\!-\!1,k) 
         \  {\rm for}  \  1\leq i \leq j \leq L,
\label{eq:recursion}
\end{eqnarray}
which arises from 
 taking away the rightmost elements of the type 
$hxxx\cdots x$ with $(j-k)$ $x$'s
from the substring of length $i+j$. 
At the end, the recurrance formula
Eq. (\ref{eq:recursion}) yields
the weighting factor $a_i$
for the substring `$xhxhx\cdots hx$' with $m$ $x$'s and $m-1$ 
$h$'s:
\begin{eqnarray}
  a_i=b(i-1,i)=b(i,i) \quad {\rm for} \quad 1 \leq i \leq L. 
\end{eqnarray}

\begin{figure}[t]
\includegraphics[width=5cm, height=5cm]{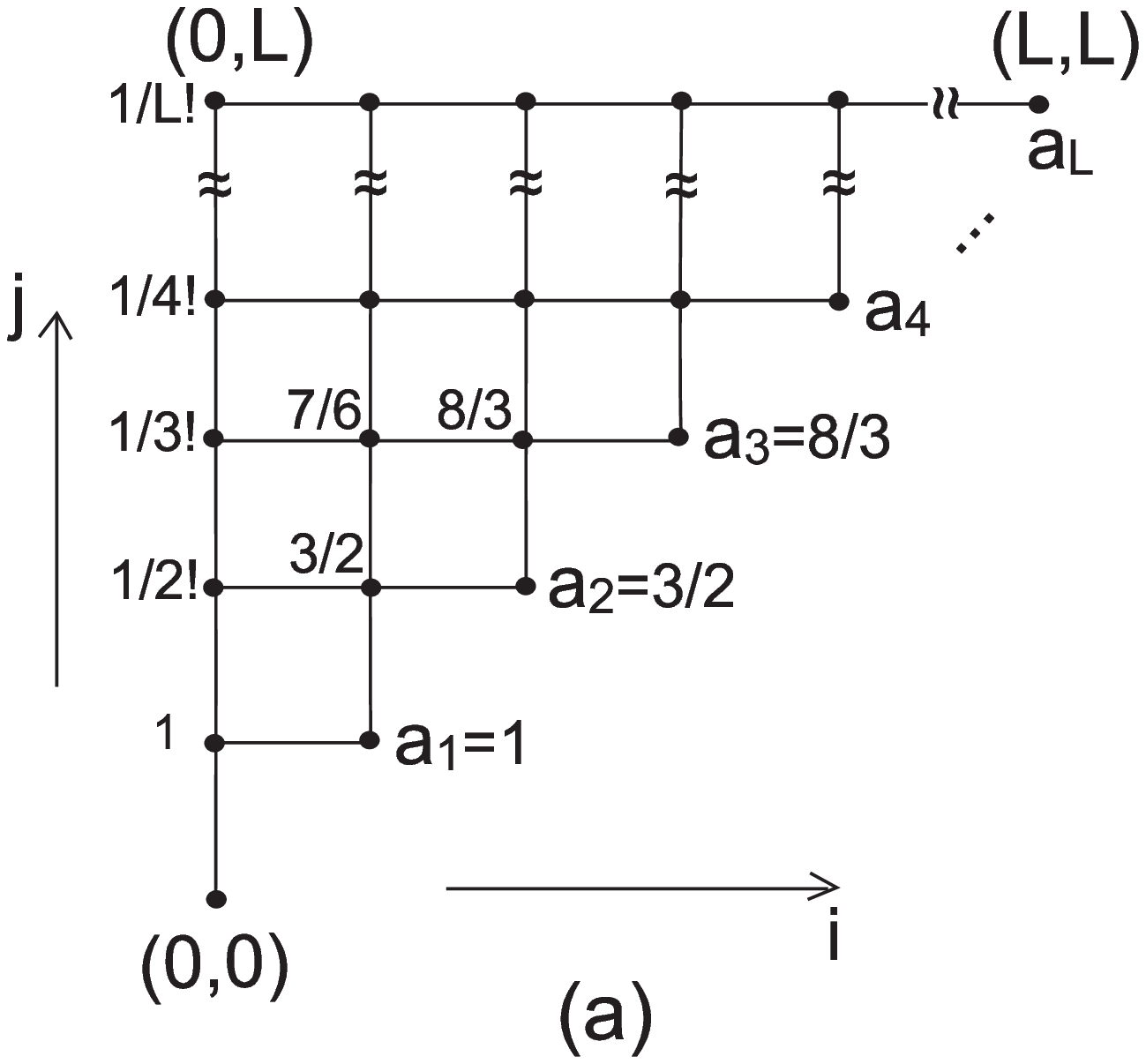}
\includegraphics[width=5cm, height=5cm]{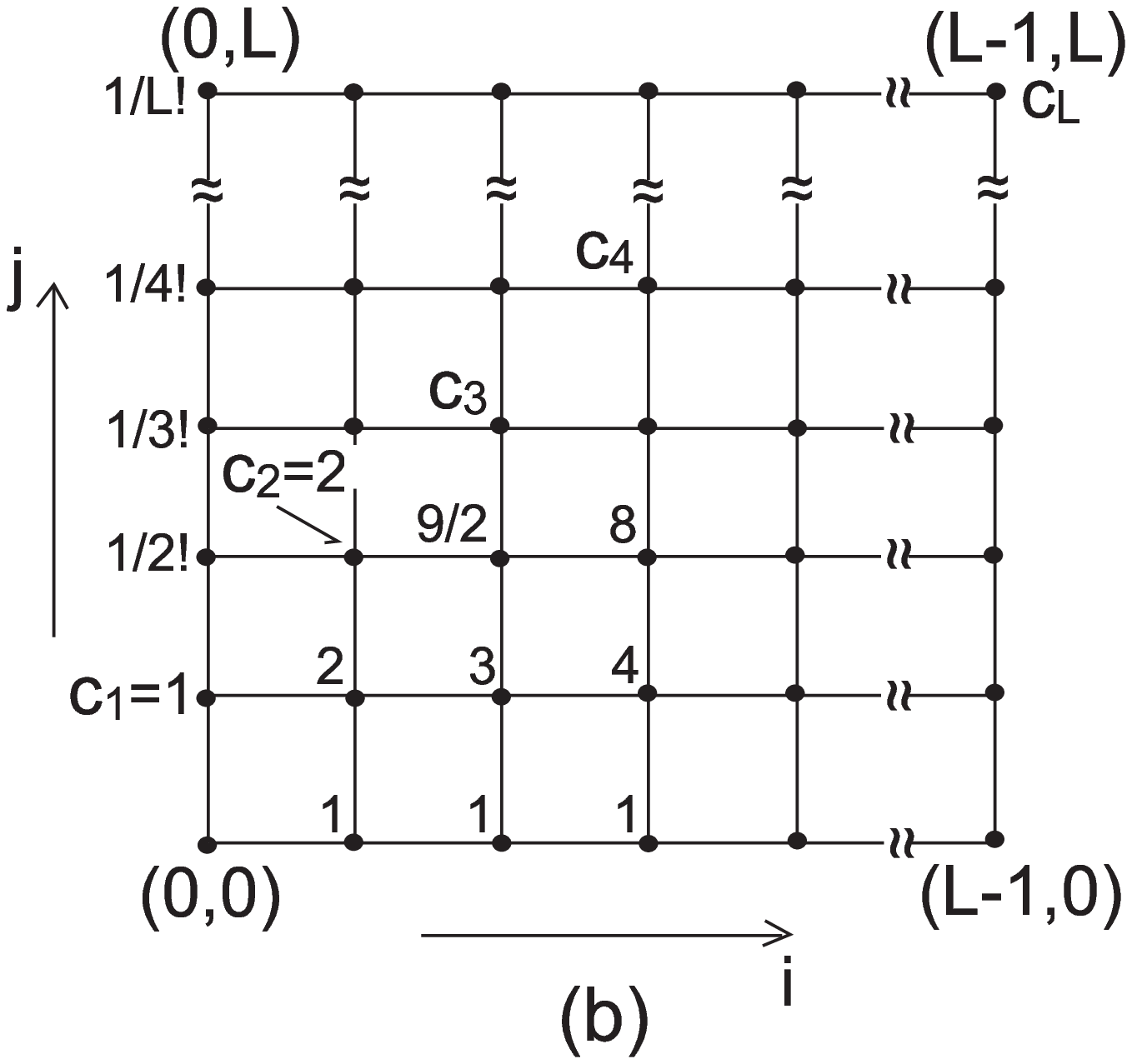}
\caption{A schematic representation of the calculation
of (a) $a_m$ and (b) $c_m$.
The coordinates in the $i$ and $j$ directions 
represent the numbers of $h$ and $x$, respectively.
The number at a point $(i,j)$ denotes (a) $b(i,j)$ and (b) $d(i,j)$,
which are recursively calculated with 
Eq.\ (\ref{eq:b}) and (\ref{eq:d}), respectively.}
\label{fig:am}
\end{figure}

Only when the last $x$ in $xhxhx...hx$ is situated at
the end of the imaginary-time interval ($L$) do 
we use the factor $c_m$ instead, which is obtained 
via the slightly different recurrence formula
[see Fig.\ \ref{fig:am}(b)],
\begin{eqnarray}\label{eq:d}
  &&d(0,j)=\frac{1}{j!} \quad {\rm for} \quad 0 \leq j \leq L, \nonumber\\
  &&d(i,j)=\sum_{k=0}^j 
           \frac{1}{(j-k)!}d(i\!-\!1,k) \nonumber \\
  &&\quad {\rm for} \quad 1\leq i \leq L-1 \ {\rm and}\  
    0 \leq j \leq L,\nonumber\\
  &&c_i=d(i-1,i) \quad {\rm for} \quad  1 \leq i \leq L. 
\end{eqnarray}

From the $a$'s and $c$'s, the total weight $F$ is calculated by multiplying
the contributions ${a_m}$ and ${c_m}$ from each 
$xhxhx...hx$-type substring in the Boltzmann factor
and the volume $L^{-k}$.
For example, for $L=8$,
\begin{eqnarray}
  &&F(2;1,1,0,0,0,0,0,0)=a_2 L^{-2},\nonumber\\
  &&F(5;0,1,0,1,0,1,1,1)=c_3 L^{-5},\nonumber\\
  &&F(6;1,0,1,1,0,1,1,1)=a_2 c_3 L^{-6}.
\end{eqnarray}
In the first example, the Boltzmann factor is $hxhxhhhhhh$.
This array replaces itself and $hxxhhhhhhh$, which
is weighted $\frac{1}{2!}$, 
and thereby the factor is $a_2=1+\frac{1}{2!}=\frac{3}{2}$
multiplied by $L^{-2}$.
The second example corresponds to $hhxhhxhhxhxhx$,
where the substring to be multiplied by a factor 
is only the last part $xhxhx$.
Therefore the factor is $c_3 L^{-5}$.
In the last example $hxhhxhxhhxhxhx$, two substrings
$xhx$ and $xhxhx$ are multiplied by $a_2$ and 
$c_3$, respectively.
So the total factor is $a_2 \times c_3 L^{-6}$.

\begin{references}
\item[\nonumber]
$^*$ Present address: Condensed Matter Theory Laboratory, RIKEN, Wako, Saitama 351-0198, Japan.
\bibitem{s36}
    J. C. Slater, Phys. Rev. \textbf{49}, 537 (1936).
\bibitem{if98}
    M. Imada, A. Fujimori, and Y. Tokura, 
     Rev. Mod. Phys. \textbf{70}, 1039 (1998);
    M. J. Rozenberg, Phys. Rev. B \textbf{55}, R4855 (1997);
    J. E. Han, M. Jarrell, and D. L. Cox, 
     {\it ibid}. \textbf{58}, R4199 (1998);
    Y. ${\rm \bar{O}}$no, R. Bulla, and M. Potthoff, 
     {\it ibid}. \textbf{67}, 035119 (2003).
\bibitem{pb05}
    Th. Pruschke and R. Bulla, Eur. Phys. J. B \textbf{44}, 217 (2005).
\bibitem{an02}
    V. I. Anisimov, I. A. Nekrasov, D. E. Kondakov, T. M. Rice, and M. Sigrist,
     Eur. Phys. J. B \textbf{25}, 191 (2002)
\bibitem{l03}
    A. Liebsch, 
     Phys. Rev. Lett. \textbf{91}, 226401 (2003);
     Phys. Rev. B \textbf{70}, 165103 (2004)
\bibitem{kk04}
    A. Koga, N. Kawakami, T. M. Rice, and M. Sigrist, Phys. Rev. Lett. 
     \textbf{92}, 216402 (2004).
\bibitem{ik05}
    K. Inaba, A. Koga, S. Suga, and N. Kawakami, 
     J. Phys. Soc. Jpn. \textbf{74}, 2393 (2005); 
     Phys. Rev. B \textbf{72}, 085112 (2005);
    M. Ferrero, F. Becca, M. Fabrizio, and M. Capone,
     {\it ibid}. \textbf{72}, 205126 (2005);
    C. Knecht, N. Bl\"{u}mer, and P. G. J. van Dongen, 
     {\it ibid}. \textbf{72}, 081103(R) (2005);
    A. Liebsch,    Phys. Rev. Lett. \textbf{95}, 116402 (2005);
    A. Koga, K. Inaba, and N.  Kawakami,
     Prog. Theor. Phys. Suppl. {\bf 160}, 253 (2005);
    S. Biermann,  L. de' Medici, and A. Georges,
     Phys. Rev. Lett. {\bf 95}, {206401}  (2005);
    P. G. J. van Dongen, C. Knecht, and N. Bl{\"u}mer,
     Phys. Status Solidi B {\bf 243}, {116} (2006);
    A. Liebsch and T. A. Costi, Euro. Phys. J. B. {\bf 51}, 523 (2006).
\bibitem{mg05}
    L. de' Medici, A. Georges, and S. Biermann, 
     Phys. Rev. B \textbf{72}, 205124 (2005)
\bibitem{kk05}
    A. Koga, N. Kawakami, T. M. Rice, and M. Sigrist, 
     Phys. Rev. B \textbf{72}, 045128 (2005).
\bibitem{ah05}
    R. Arita and K. Held, Phys. Rev. B \textbf{72}, 201102(R) (2005).
   
\bibitem{t00}
   T. Takimoto, Phys. Rev. B \textbf{62}, R14641 (2000);
   M. Capone, M. Fabrizio, C. Castellani, and E. Tosatti, 
    Science \textbf{296}, 2364 (2002);
   T. Hotta and K. Ueda, Phys. Rev. Lett. \textbf{92}, 107007 (2004);
   Y. Yanase, M, Mochizuki, and M. Ogata, 
    J. Phys. Soc. Jpn. \textbf{74}, 430 (2005);
   M. Mochizuki, Y. Yanase, and M. Ogata, 
    Phys. Rev. Lett. \textbf{94}, 147005 (2005).
\bibitem{h04}
An alternative, continuous auxiliary field has been 
proposed by J. E. Han, Phys. Rev. B \textbf{70}, 054513 (2004).
\bibitem{sa04}
   S. Sakai, R. Arita, and H. Aoki, Phys. Rev. B \textbf{70}, 172504 (2004).
\bibitem{mv89}
   W. Metzner and D. Vollhardt, Phys. Rev. Lett. \textbf{62}, 324 (1989).
\bibitem{gk96}
   A. Georges, G. Kotliar, W. Krauth, and M. J. Rozenberg,
    Rev. Mod. Phys. \textbf{68}, 13 (1996).
\bibitem{ap97}
   V. I. Anisimov, A. I. Poteryaev, M. A. Korotin, A. O. Anokhin, and 
   G. Kotliar, J. Phys.: Condens. Matter \textbf{9}, 7359 (1997);
A.I. Lichtenstein and M.I. Katsnelson, Phys. Rev. B
 {\bf 57}, {6884} (1998).

\bibitem{ks05}
   G. Kotliar, S. Y. Savrasov, K. Haule, V. S. Oudovenko, O. Parcollet,
    and C.A. Marianetti, Rev. Mod. Phys. \textbf{78}, 865 (2006);
   K. Held, cond-mat/0511293 (unpublished).
\bibitem{ck94}
   M. Caffarel and W. Krauth, Phys. Rev. Lett. \textbf{72}, 1545 (1994).
\bibitem{hf85}
   J. E. Hirsch and R. M. Fye, Phys. Rev. Lett. \textbf{56}, 2521 (1986).
\bibitem{cm05}
   M. Capone, L. de'Medici, and A. Georges, cond-mat/0512484 (unpublished).

\bibitem{mm03}
   A. P. Mackenzie and Y. Maeno, Rev. Mod. Phys. \textbf{75}, 657 (2003).


\bibitem{h83}
   J. E. Hirsch, Phys. Rev. B \textbf{28}, 4059 (1983); 
   \textbf{29}, 4159 (1984).

\bibitem{remark1}
   Exceptions are Refs.\ \onlinecite{h04, sa04, kk05}, and \onlinecite{ah05}.
\bibitem{hv98}
   K. Held and D. Vollhardt, Eur. Phys. J. B \textbf{5}, 473 (1998).
\bibitem{remark2}
   Although the decomposition to each two-orbital part is allowed 
   within an accuracy $\sim O(\D\t^2)$,
   it breaks the equivalence of the interorbital interactions and 
   will also cause a severe sign problem.
\bibitem{rh99}
   S. M. A. Rombouts, K. Heyde, and N. Jachowicz, Phys. Rev. Lett. 
   \textbf{82}, 4155 (1999).
\bibitem{bs81}
   R. Blankenbecler, D. J. Scalapino, and R. L. Sugar, Phys. Rev. D 
   \textbf{24}, 2278 (1981).
\bibitem{rl04}
   A. N. Rubtsov, V. V. Savkin, and A. I. Lichtenstein, 
    Phys. Rev. B \textbf{72}, 035122 (2005).
\bibitem{wc05}
   P. Werner, A. Comanac, L. de' Medici, A. J. Millis, and M. Troyer,
   Phys. Rev. Lett. {\bf 97}, 076405 (2006).
\bibitem{sa06}
   S. Sakai, R. Arita, and H. Aoki, Physica B {\bf 378-380}, 288 (2006).
\bibitem{remark3}
   Although $\hat{H}_0$ is formally expressed as the one-body part of 
   the multiorbital Anderson Hamiltonian, the explicit form is not 
   necessary in the following since the QMC algorithm only uses the 
   relation between the Green functions.
\bibitem{rh98}
   S. M. A. Rombouts, K. Heyde, and N. Jachowicz, Phys. Lett. A
   \textbf{242}, 271 (1998).
\bibitem{remark4}
   In practice we can further reduce the number of auxiliary fields:
   $t_\ua$ and $t_\da$ in Eq.\ (\ref{eq:HSJ3}) are not necessary
   when we combine these terms with $U'-J$ terms 
   in Eq.\ (\ref{HSU}) to decouple them simultaneously.

\bibitem{hv00} 
   K. Held and D. Vollhardt,
     Phys. Rev. Lett. {\bf 84}, 5168 (2000).
\bibitem{lk01}
   A. I. Lichtenstein, M. I. Katsnelson, and G. Kotliar, 
     Phys. Rev. Lett. {\bf 87}, 67205 (2001).
\bibitem{ll00}
   A. Liebsch and A. Lichtenstein, 
     Phys. Rev. Lett. \textbf{84}, 1591 (2000).
\bibitem{cm66}
   A. Callaghan, C. W. Moeller and R. Ward,
     Inorg. Chem. \textbf{5}, 1572 (1966).
\bibitem{pn06}
   Z. V. Pchelkina, I. A. Nekrasov, Th. Pruschke, 
   A. Sekiyama, S. Suga, V. I. Anisimov, and D. Vollhardt,
     cond-mat/0601507 (unpublished).
\bibitem{fh04}
   M. Feldbacher, K. Held, and F. F. Assaad, 
    Phys. Rev. Lett. \textbf{93}, 136405 (2004).

\end{references}
\end{document}